\documentclass[preprint,aps,amsmath,amssymb,nofootinbib,tightenlines,superscriptaddress]{revtex4}

\usepackage{graphicx}
\usepackage{axodraw}
\usepackage{dcolumn}
\usepackage{bm}

\newcommand{\be}{\begin{equation}}
\newcommand{\ee}{\end{equation}}
\newcommand{\bea}{\begin{eqnarray}}
\newcommand{\eea}{\end{eqnarray}}

\newcommand{\bay}{\begin{array}}
\newcommand{\eay}{\end{array}}

\newcommand{\nln}{\nonumber}

\makeatletter       

\renewcommand{\p@subsection}{}

\renewcommand{\p@subsubsection}{}

\makeatother

\begin{document}


\title{QCD Corrections to $K-\bar{K}$ Mixing \\ in R-symmetric Supersymmetric Models}

\author{Andrew E. Blechman}%
\email{blechman@physics.utoronto.ca}%
\affiliation{Department of Physics, University of Toronto,
Toronto, Ontario M5S 1A7, Canada.}

\author{Siew-Phang Ng}%
\email{spng@physics.utoronto.ca}%
\affiliation{Department of Physics, University of Toronto,
Toronto, Ontario M5S 1A7, Canada.}

\date{\today  \\ \vspace{1in}}

\begin{abstract}\vspace{0.1in}
The leading-log QCD corrections to $K-\bar{K}$ mixing in R--symmetric supersymmetric models are computed using effective field theory techniques.  The spectrum topology where the gluino is significantly heavier than the squarks is motivated and focused on.  It is found that, like in the MSSM, QCD corrections can tighten the kaon mass difference bound by roughly a factor of three.  CP violation is also briefly considered, where QCD corrections can constrain phases to be as much as a factor of ten smaller than the uncorrected value.
\end{abstract}

\maketitle

\section{Introduction}

It has recently been proposed in \cite{KPW} that by extending the usual minimal supersymmetric standard model (MSSM) to a theory with an additional $U(1)_R$ symmetry, the SUSY flavor problem can be solved, at least in part.  In particular, the additional $R$ symmetry forbids several flavor-changing neutral current (FCNC) operators usually found in the low energy effective theory of the MSSM; and those operators that are still there can be suppressed in large regions of parameter space, even for sizable flavor mixing angles.  This is in contrast to the usual requirement of a ``flavor-blind" mechanism, such as gauge mediation, to forbid the presence of such mixing angles.  

Models with a (partial) $U(1)_R$ symmetry were considered several years back in \cite{Polchinski:1982an,Hall:1990hq,Fox:2002bu}, where the gauginos were given a Dirac mass.  Several papers have extended on this idea over the years \cite{BKW,p0,p1,p2,p3,p4,p5,p6,p7,p8,p9,p10}.  What was shown in \cite{KPW} was that in regions of parameter space where the gluino is much heavier than the squarks, the FCNCs are under control even for sizable mixing angles.  Fortunately, it seems that such spectrum topologies are rather generic in models with Dirac gauginos \cite{BKW}.

Of course, a viable model of supersymmetry cannot have a completely unbroken $R$ symmetry, since such a symmetry is unavoidably broken by the gravitino mass which arises due to the Super-Higgs mechanism whenever you break supersymmetry.  This will inevitably generate $R$ symmetry breaking in the low-energy sector via the Giudice-Masiero mechanism or anomaly mediation or some such device.  However, one can imagine that such effects are somehow hidden or rendered irrelevant by an as-yet-unknown UV completion.  Some thoughts along these lines were presented in \cite{KPW}.

Even with these criticisms, the prospect of being able to solve the SUSY flavor problem without resorting to the usual flavor-blind requirements is exciting, to say the least.  In \cite{KPW}, a tree-level analysis of flavor changing effects from gluino box diagrams was considered.  In this note, we compute the one-loop QCD corrections to these results.  This is believed to be the most important correction to $K-\bar{K}$ mixing, although there are scenarios where electroweak box diagrams can also be important.  We will follow closely the outline of \cite{Bagger:1997gg}, where a similar analysis was done for the usual MSSM.

\section{The Effective Lagrangian}

The $R$ symmetry can be seen to give some very powerful constraints on the low-energy MSSM:

\begin{itemize}
\item No Majorana masses for the gauginos.
\item No A-terms among the scalar fields.
\item No $\mu$ term for the Higgs, although there is a $B_\mu$ term.
\end{itemize}

Putting these facts together leads right away to the result that the only coupling between the ``left-handed" squarks and the ``right-handed" squarks (and sleptons) is through the usual Yukawa interactions, and these are flavor diagonal.  Therefore, any flavor-changing couplings can only come from the soft squark mass matrices, and these are necessarily of the form ``left-left" or ``right-right". In what follows, we will assume that the squarks are nearly degenerate in mass (this is a reasonable assumption for the first two generations in an mSUGRA-like scenario), and parametrize the off-diagonal elements of the mass matrix in terms of dimensionless numbers\footnote{These quantities are usually referred to as $\delta_{LL},\delta_{RR}$ in the literature, but since there is no flavor mixing of the ``left-right" form, we drop the second $L(R)$ for notational simplicity.}:
\be
\delta_L\equiv \frac{m_{\tilde{Q}12}^2}{M_{\tilde{q}}^2}\qquad
\delta_R\equiv \frac{m_{\tilde{d}12}^2}{M_{\tilde{q}}^2}
\label{qmix}
\ee
where $\tilde{Q}$ is the left-handed squark and $\tilde{d}$ is the right-handed down-type squark and $M_{\tilde{q}}$ is the universal squark mass.  These are the only relevant squark-mixing angles for $K-\bar{K}$ mixing.

To compute the amplitudes in a process with several scales, the best way to proceed is through the construction of an effective theory, where we construct operators of lowest possible dimension using only the low energy degrees of freedom, in this case the quarks. To this end, we can write down an effective Lagrangian of the form
\be
\mathcal{L}_{\rm eff}=\frac{\alpha_s^2(M_{\tilde{g}})}{216}\left(\frac{M_{\tilde{q}}^2}{M_{\tilde{g}}^4}\right)\sum_{n}C_n(\mu)\mathcal{O}_n(\mu) \label{Leff}
\ee
where the leading operators\footnote{In this list the fermions are 4-component fields, whereas in the rest of this paper, we use 2-component notation.  It is very easy to go from one to the other: just replace $\psi_R\leftrightarrow\bar{\psi}_R$ and leave the $\psi_L$ fields alone.} are dimension 6:
\bea
\mathcal{O}_1&=&(\bar{d}^i_L\gamma^\mu s^i_L)(\bar{d}^j_L\gamma_\mu s^j_L) \label{op1} \\
\mathcal{O}_4&=&(\bar{d}^i_R s^i_L)(\bar{d}^j_L s^j_R) \label{op4} \\
\mathcal{O}_5&=&(\bar{d}^i_R s^j_L)(\bar{d}^j_L s^i_R) \label{op5}
\eea
$i,j$ are color indices, and there is a similar operator $\tilde{\mathcal{O}}_1$ with $L\leftrightarrow R$.  We are using the same numbering scheme and normalization as \cite{Bagger:1997gg} which was also used in \cite{KPW}.  Notice that the lack of any ``left-right" couplings means that $C_{2,3}=0$ in their operator list.  The $\mu$ in Equation \ref{Leff} is the renormalization scale, the dependence of which cancels between the operator and the Wilson coefficient.  This implies that the Wilson coefficients must obey the Renormalization Group Equation:
\be
\mu\frac{dC_m}{d\mu}=\gamma^T_{mn}C_{n}
\ee
where $\gamma_{mn}\equiv(Z^{-1})_{ma}\frac{dZ_{an}}{d\log\mu}$ is the anomalous dimension matrix.

There are three spectrum topologies that one can imagine: (1) $M_{\tilde{q}}\sim M_{\tilde{g}}$, (2) $M_{\tilde{q}}\gg M_{\tilde{g}}$ and (3) $M_{\tilde{q}}\ll M_{\tilde{g}}$.  The first two cases turn out to be identical to the usual MSSM done in \cite{Bagger:1997gg,SUSYFCNC}, with $\delta_{LR}=f_6(x)=0$.  However, case (3) turns out to be quite different.  Since these models are most favorable in this case, we will present this calculation in detail.

We proceed in four stages.  The first step is to integrate out the gluinos at the scale $M_{\tilde{g}}$, where we match to an intermediate theory with quarks and squarks as the degrees of freedom.  We will then run the theory down to the scale $M_{\tilde{q}}$, which allows us to resum the leading-log QCD corrections between the two scales.  Next, we match our intermediate theory to Equation \ref{Leff}.  Finally, we can do the usual QCD running of this theory down to the hadronic scale, integrating out the heavy quarks as we go.  Ideally we would like to run the theory down to the mass of the $K$ meson, but QCD is strongly coupled at such scales, so we will stop the running at a scale $\mu_{\rm had}\sim 1$ GeV defined by the condition that $\alpha_s(\mu_{\rm had})=1$.


\subsection{Step 1: Matching to the Intermediate Theory}

\subsubsection{General Operator Analysis}

\begin{figure}[tb]
\centerline{\scalebox{0.7}{\includegraphics{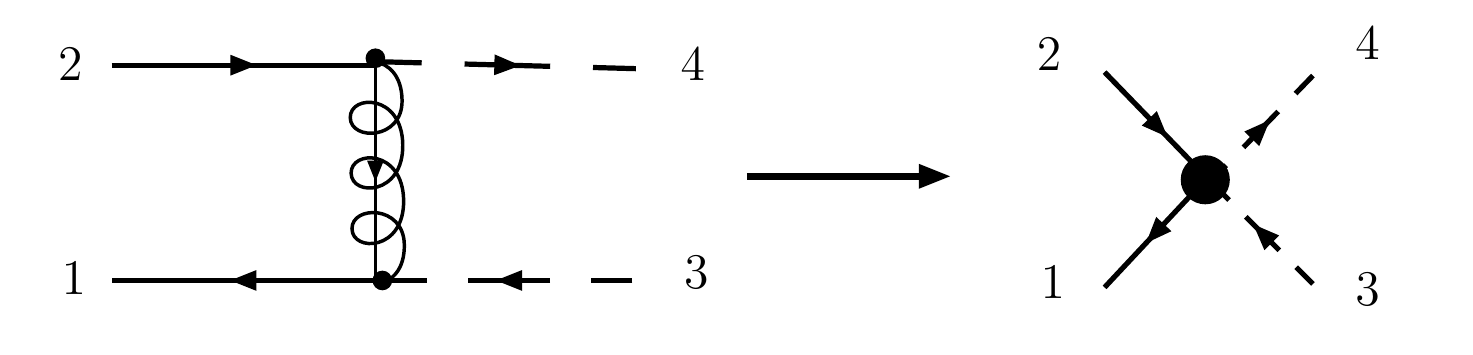}}}
\vspace{-0.2cm}
\caption{Integrating out the gluino.  In this and all Feynman diagrams, momentum is flowing left to right, and the arrows on the fermion lines represent spinor flow.}
\label{sqcd2Q}
\end{figure}

The gluino couplings to the quark-squark are given in the full theory by
\be
\Delta\mathcal{L}=-\sqrt{2}g_s[(\tilde{q}^*T^aq)\lambda^a + \bar{\lambda}^a(\bar{q}T^a\tilde{q})]
\ee
Here $T^a$ are the $SU(3)$ generators in the fundamental representation, and $\lambda^a$ is the Weyl spinor field representing the part of the gluino that interacts directly with matter.  Since it no longer has a Majorana mass, the only propagators are of the form $\langle\bar{\lambda}\lambda\rangle$ and the diagram in Figure \ref{sqcd2Q} gives
\be
i\mathcal{M}=-2g_s^2T^a_{ij}T^a_{mn}\left[(\bar{q}_1^i\tilde{q}_3^j)\frac{-i\bar{\sigma}\cdot(p_1-p_3)}{(p_1-p_3)^2-M_{\tilde{g}}^2}(\tilde{q}_4^{m*}q_2^n)\right]
\ee
This amplitude then matches onto an intermediate effective theory with six operators up to flavor structure:
\be
\mathcal{L}_{\rm int}=\frac{g_s^2(M_{\tilde{g}})}{M_{\tilde{g}}^2}\sum_{i=1}^6 D_i(\mu)\mathcal{Q}_i(\mu) \label{Lint}
\ee
where
\bea
\mathcal{Q}_1&=&[(i\mathcal{D}_\mu\bar{q}_1)^i\bar{\sigma}^\mu q_2^i](\tilde{q}_3^j\tilde{q}_4^{j*}) +[\bar{q}_1^i\bar{\sigma}^\mu q_2^i]((i\mathcal{D}_\mu\tilde{q}_3)^j\tilde{q}_4^{j*})  \label{Qop1}\\
\mathcal{Q}_2&=&[(i\mathcal{D}_\mu\bar{q}_1)^i\bar{\sigma}^\mu q_2^j](\tilde{q}_3^i\tilde{q}_4^{j*}) +[\bar{q}_1^i\bar{\sigma}^\mu q_2^j]((i\mathcal{D}_\mu\tilde{q}_3)^i\tilde{q}_4^{j*}) \label{Qop2} \\
\mathcal{Q}_3&=&[(i\mathcal{D}_\mu\bar{q}_1)^i\bar{\sigma}^\mu q_2^i](\tilde{q}_3^j\tilde{q}_4^{j*}) +[\bar{q}_1^i\bar{\sigma}^\mu (i\mathcal{D}_\mu q_2)^i](\tilde{q}_3^j\tilde{q}_4^{j*})  \label{Qop3} \\
\mathcal{Q}_4&=&[(i\mathcal{D}_\mu\bar{q}_1)^i\bar{\sigma}^\mu q_2^j](\tilde{q}_3^i\tilde{q}_4^{j*}) +[\bar{q}_1^i\bar{\sigma}^\mu (i\mathcal{D}_\mu q_2)^j](\tilde{q}_3^i\tilde{q}_4^{j*})  \label{Qop4} \\
\mathcal{Q}_5&=&[(i\mathcal{D}_\mu\bar{q}_1)^i\bar{\sigma}^\mu q_2^i](\tilde{q}_3^j\tilde{q}_4^{j*}) +[\bar{q}_1^i\bar{\sigma}^\mu q_2^i](\tilde{q}_3^j(i\mathcal{D}_\mu\tilde{q}^{*}_4)^{j}) \label{Qop5} \\
\mathcal{Q}_6&=&[(i\mathcal{D}_\mu\bar{q}_1)^i\bar{\sigma}^\mu q_2^j](\tilde{q}_3^i\tilde{q}_4^{j*}) +[\bar{q}_1^i\bar{\sigma}^\mu q_2^j](\tilde{q}_3^i(i\mathcal{D}_\mu\tilde{q}^{*}_4)^{j})  \label{Qop6}
\eea
Here, $\mathcal{D}_\mu$ is the QCD covariant derivative required by gauge invariance, and each operator has a Wilson coefficient:
\bea
D_1(M_{\tilde{g}})&=&-1+\mathcal{O}(g_s^2) \label{D1} \\
D_2(M_{\tilde{g}})&=&+\frac{1}{3} +\mathcal{O}(g_s^2) \label{D2} \\
D_{3-6}(M_{\tilde{g}})&=&\mathcal{O}(g_s^2) \label{D36} 
\eea

Note that unlike the situation in \cite{Bagger:1997gg} these operators are dimension 6, not dimension 5.  This point was emphasized in \cite{KPW} and stems from the key fact that there is no gluino Majorana mass, and hence no $\langle\lambda\lambda\rangle$ propogator.

\subsubsection{Class A versus Class B Operators}

In order to match the intermediate theory to Equation \ref{Leff} we must specify the flavor structure of the operators.  There are four flavor combinations that we will need, and these can be divided into two classes of operators depending on their color structure.  The ``Class A" operators corresponding to Equation \ref{Qop1} are:
\bea
\mathcal{Q}_1^1&=&[(i\mathcal{D}_\mu\bar{d}_L)^i\bar{\sigma}^\mu s_L^i](\tilde{d}_L^j\tilde{s}_L^{j*}) +[\bar{d}_L^i\bar{\sigma}^\mu s_L^i]((i\mathcal{D}_\mu\tilde{d}_L)^j\tilde{s}_L^{j*}) \label{Qopfl1}\\
\mathcal{Q}_1^2&=&[(i\mathcal{D}_\mu\bar{d}_R)^i\bar{\sigma}^\mu s_R^i](\tilde{d}_R^j\tilde{s}_R^{j*}) +[\bar{d}_R^i\bar{\sigma}^\mu s_R^i]((i\mathcal{D}_\mu\tilde{d}_R)^j\tilde{s}_R^{j*}) \label{Qopfl2}
\eea
and the ``Class B" operators are:
\bea
\mathcal{Q}_1^3&=&[(i\mathcal{D}_\mu\bar{d}_L)^i\bar{\sigma}^\mu d_R^j](\tilde{d}_L^j\tilde{d}_R^{i*}) +[\bar{d}_L^i\bar{\sigma}^\mu d_R^j]((i\mathcal{D}_\mu\tilde{d}_L)^j\tilde{d}_R^{i*}) \label{Qopfl3}\\
\mathcal{Q}_1^4&=&[(i\mathcal{D}_\mu\bar{s}_R)^i\bar{\sigma}^\mu s_L^j](\tilde{s}_R^j\tilde{s}_L^{i*}) +[\bar{s}_R^i\bar{\sigma}^\mu s_L^j]((i\mathcal{D}_\mu\tilde{s}_R)^j\tilde{s}_L^{i*}) \label{Qopfl4}
\eea
and similarly for $\mathcal{Q}^I_{2-6}$.  Since QCD is vector-like and flavor blind, all of these operators will share the same matching coefficients $D_i(M_{\tilde{g}})$.  However, since the color flow is different\footnote{This is only true for $\mathcal{Q}_{1,3,5}$; for $\mathcal{Q}_{2,4,6}$ the color flow is identical between Class A and Class B operators.}, the Class A operators will run differently than the Class B operators: see the appendix for details.

\subsection{Step 2: Running the Intermediate Theory}

To compute the one-loop running of the operators $\mathcal{Q}_{i}$ we must evaluate the diagrams shown in Figure \ref{running}.  Performing this calculation is a straightforward (if tedious) exercise and yields the result for the anomalous dimension matrix presented in the appendix.

Now we can compute the Wilson coefficients at the squark mass scale, using $b_0=-5$
\be
D^I_i(M_{\tilde{q}})=\left[\eta^{-\tilde{\gamma}_I^T/5}\right]_{ij}D^I_j(M_{\tilde{g}})
\ee
where we define
\be
\eta=\frac{\alpha_s(M_{\tilde{q}})}{\alpha_s(M_{\tilde{g}})}
\ee
and $\tilde{\gamma}$ is the anomalous dimension matrix without the $\alpha_s/2\pi$ prefactor.  The index $I$ is for whether the operator is a Class A or a Class B operator.  The solutions are quite messy and are computed numerically in what follows.

\subsection{Step 3: Matching to the Low Energy Effective Theory}

\begin{figure}[tb]
\centerline{\scalebox{0.7}{\includegraphics{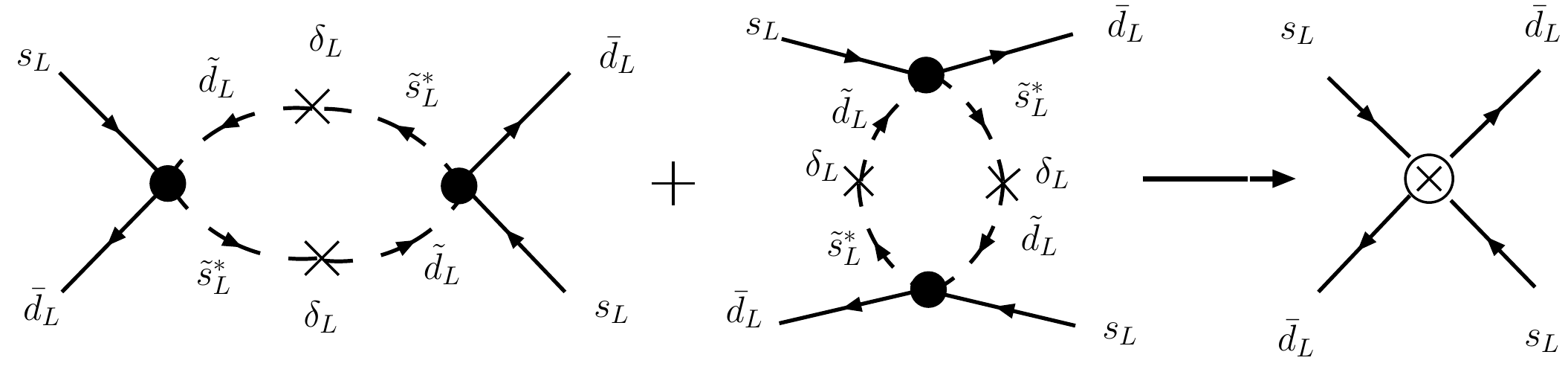}}}
\vspace{-0.2cm}
\caption{Matching calculation to $\mathcal{O}_1$ involving Class A operators.  There is an analogous diagram for $\tilde{\mathcal{O}_1}$ with $L\leftrightarrow R$.}
\label{Q2O1}
\end{figure}

\begin{figure}[tb]
\centerline{\scalebox{0.7}{\includegraphics{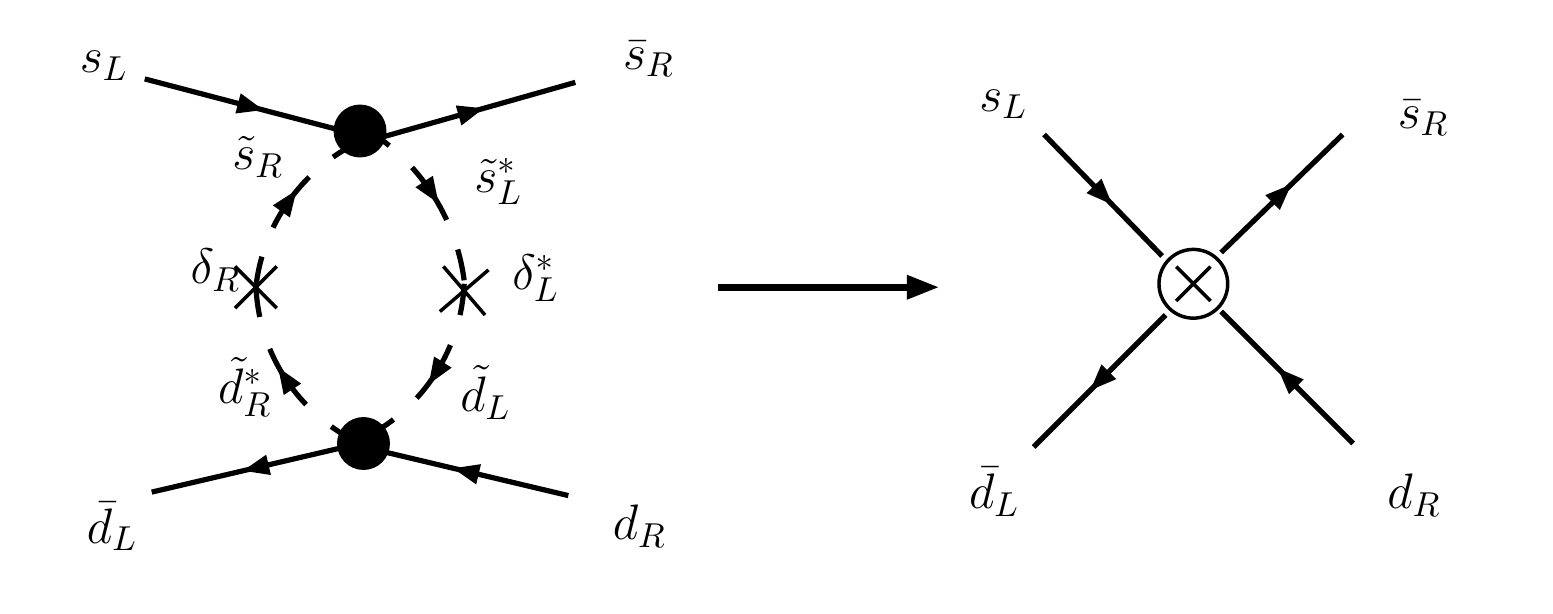}}}
\vspace{-0.2cm}
\caption{Matching calculation to $\mathcal{O}_{4,5}$ involving Class B operators.}
\label{Q2O45}
\end{figure}

We can now use the renormalization group to evolve the operators from the gluino mass scale down to the squark mass scale, and then consider the diagram where two operator insertions and two flavor-violating couplings generate the dimension 6 operators in Equation \ref{Leff}; see Figure \ref{Q2O1}-\ref{Q2O45}.  This matching will proceed as:
\bea
C_1(M_{\tilde{q}})\langle\mathcal{O}_1(M_{\tilde{q}})\rangle&=&\frac{216}{M_{\tilde{q}}^2}\sum_{i,j} \left(D^A_i(M_{\tilde{q}})D^A_j(M_{\tilde{q}})\right) \langle\mathcal{Q}^1_i(M_{\tilde{q}})\mathcal{Q}^1_j(M_{\tilde{q}})\rangle \\
\tilde{C}_1(M_{\tilde{q}})\langle\mathcal{\tilde{O}}_1(M_{\tilde{q}})\rangle&=&\frac{216}{M_{\tilde{q}}^2} \sum_{i,j} \left(D^A_i(M_{\tilde{q}})D^A_j(M_{\tilde{q}})\right) \langle\mathcal{Q}^2_i(M_{\tilde{q}})\mathcal{Q}^2_j(M_{\tilde{q}})\rangle \\
C_{4,5}(M_{\tilde{q}})\langle\mathcal{O}_{4,5}(M_{\tilde{q}})\rangle&=&\frac{216}{M_{\tilde{q}}^2}\sum_{i,j} \left(D^B_i(M_{\tilde{q}})D^B_j(M_{\tilde{q}})\right) \langle\mathcal{Q}^3_i(M_{\tilde{q}})\mathcal{Q}^4_j(M_{\tilde{q}})\rangle 
\eea
Since the $\mathcal{O}_i$ do not contain derivatives we can safely match matrix elements involving  states of quarks with vanishing momenta, where we immediately see that $\langle\mathcal{Q}_{3,4}\rangle=0$.  So these operators do not contribute in the matching, although they do contribute to the running.  By doing the loop we calculate the low energy Wilson coefficients at $\mu=M_{\tilde{q}}$ to be
%
%
\bea
C_1(M_{\tilde{q}})&=&-9\delta_L^2((3D_1^2 +2D_1D_2 +D_2^2+3D_5^2 +2D_5D_6 +D_6^2)\nln \\
& &\qquad-2(3D_1D_5+D_1D_6+D_2D_5+D_2D_6))\left.\right|_{\mu=M_{\tilde{q}}}  \label{c1} \\
C_4(M_{\tilde{q}})&=&-72\left.\delta_L^{*}\delta_R(D_1D_2+D_5D_6-D_1D_6-D_2D_5)\right|_{\mu=M_{\tilde{q}}}  \label{c4} \\
C_5(M_{\tilde{q}})&=&-36\left.\delta_L^{*}\delta_R ((D_1^2+D_2^2+D_5^2+D_6^2)-2(D_1D_5+D_2D_6))\right|_{\mu=M_{\tilde{q}}} \label{c5}
\eea
and $\tilde{C}_1=C_1$ with $\delta_L\leftrightarrow\delta_R$.  Here it is understood that the Wilson coefficients in Equation \ref{c1} are Class A while the Wilson coefficients in Equations \ref{c4}-\ref{c5} are Class B.  Plugging in the uncorrected values for the Wilson coefficients in Equation \ref{D1} -- \ref{D36} gives agreement with \cite{KPW}.

\subsection{Step 4: Running to the Hadronic Scale}

Once we have the Wilson coefficients at the squark scale, we can run the scale down to $\mu_{\rm had}$.  To do this, we must calculate the anomalous dimensions for these operators.  This has already been done in \cite{Bagger:1997gg}:
\bea
\gamma_1&=&\frac{\alpha_s}{\pi} \\
\gamma_{ij}&=&\frac{\alpha_s}{2\pi}\left(\bay{cc}-8&0\\-3&+1\eay\right)\qquad i,j=4,5
\eea
Along the way we integrate out the quarks.  This will not generate any new operators at this order, but it will change the running of $\alpha_s$.  


Finally we must calcuate the remaining matrix elements at the hadronic scale.  They are given by
\bea
\langle K|\mathcal{O}_1(\mu_{\rm had})|\bar{K}\rangle&=&\langle K|\tilde{\mathcal{O}}_1(\mu_{\rm had})|\bar{K}\rangle=\frac{1}{3}m_Kf_K^2B_1 \\
\langle K|\mathcal{O}_4(\mu_{\rm had})|\bar{K}\rangle&=&\frac{1}{4}\left[\frac{1}{6}+\left(\frac{m_K}{m_s+m_d}\right)^2\right]m_Kf_K^2B_4 \\
\langle K|\mathcal{O}_5(\mu_{\rm had})|\bar{K}\rangle&=&\frac{1}{4}\left[\frac{1}{2}+\frac{1}{3} \left(\frac{m_K}{m_s+m_d}\right)^2\right]m_Kf_K^2B_5
\eea
where $B_i$ are the bag factors characterizing the vacuum saturation approximation.

\section{Results}

\begin{table}[tb]
\begin{center}
\begin{tabular}{ccccccccccccccccccc||cccccccccccc}
$m_d$ & & & & $m_s$ & & & & $m_K$ & & & & $f_K$ & & & & $\Delta m_K$ & & & & & &$B_1$ & & & & $B_2$ & & & & $B_3$ \\
\hline
7 MeV & & & & 95 MeV & & & & 498 MeV & & & & 160 MeV & & & & $3.48\times 10^{-12}$ MeV & & & & & &0.6 & & & & 1.03 & &  & & 0.73
\end{tabular}
\end{center}
\caption{Numerical values for the parameters in the matrix elements \cite{PDG}.}
\label{params}
\end{table}

\begin{figure}[tb]
\centerline{\scalebox{0.7}{
\includegraphics{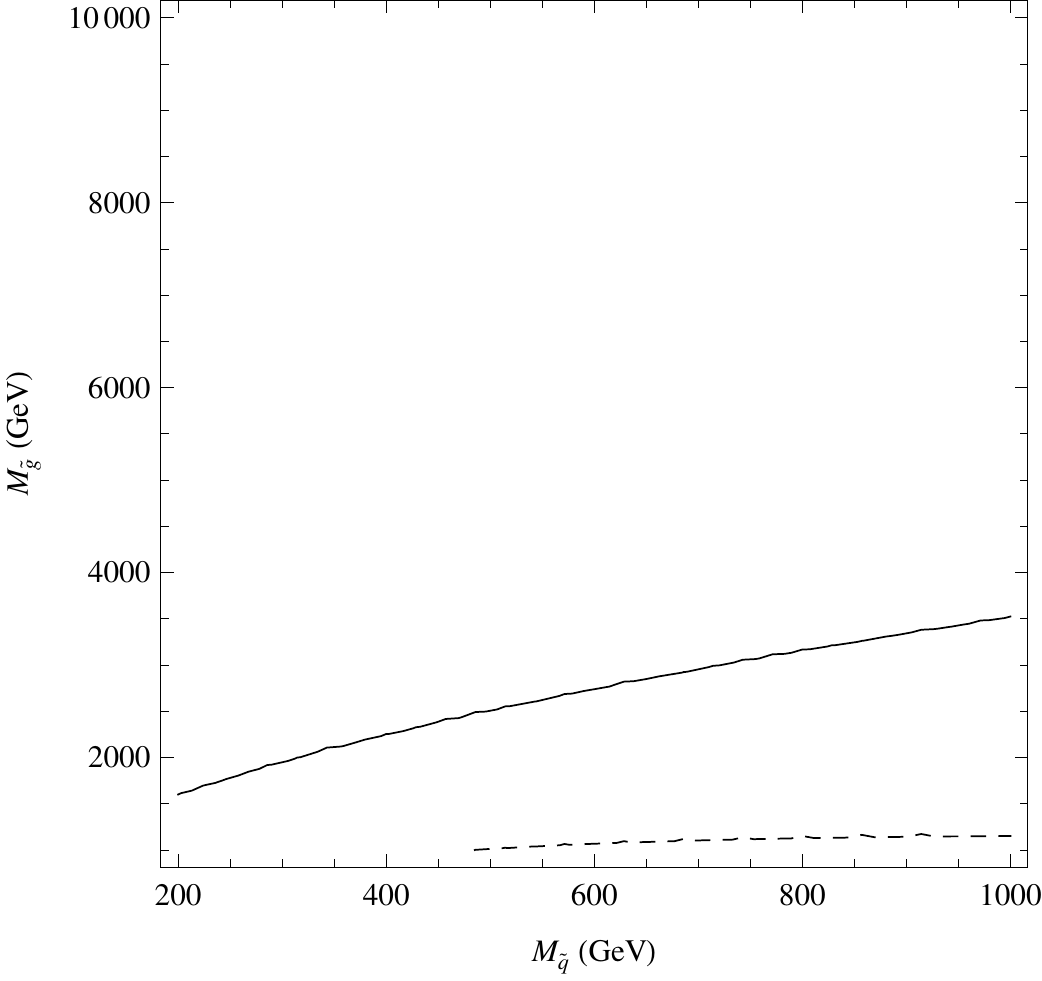}\hspace{0.15cm}
\includegraphics{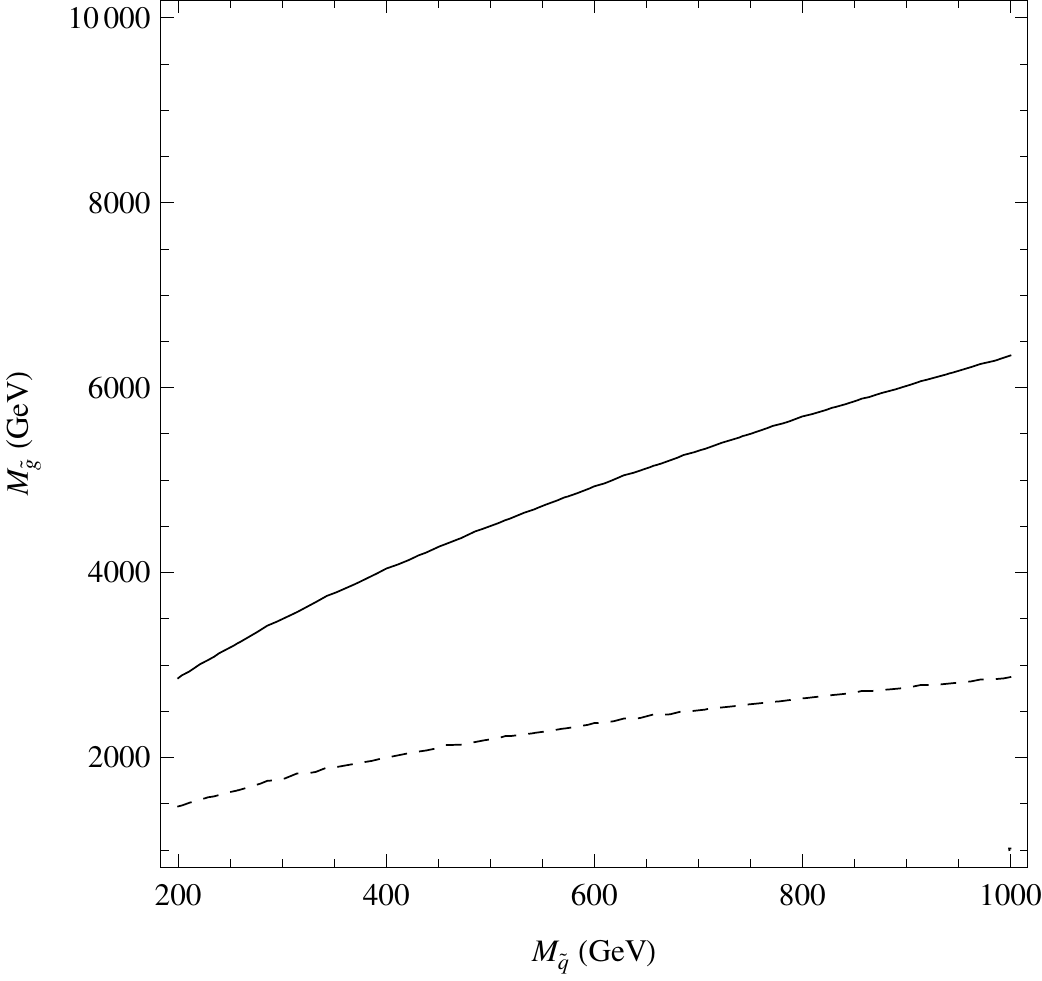}}}
\vspace{0.05cm}
\centerline{\scalebox{0.7}{
\includegraphics{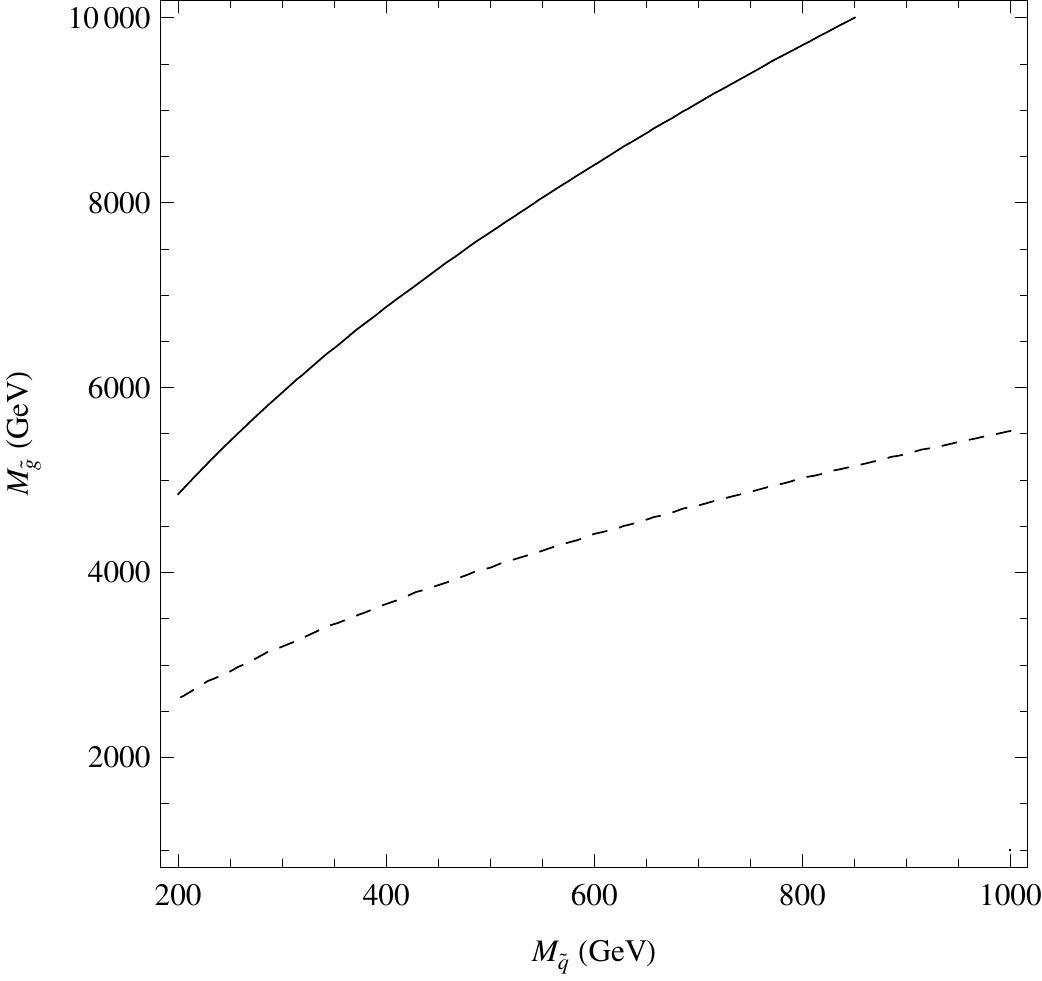}\hspace{0.15cm}
\includegraphics{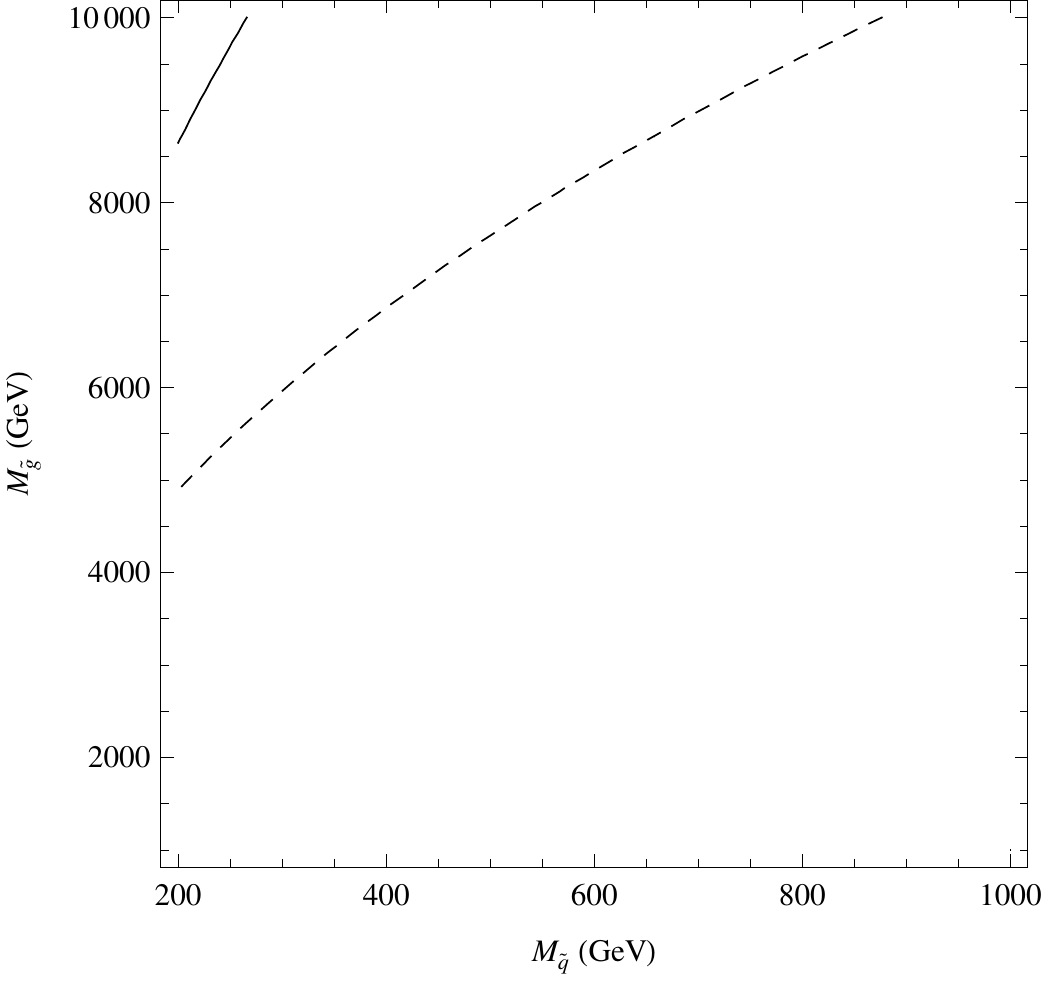}}}
\vspace{-0.2cm}
\caption{Exclusion plots in the $M_{\tilde{g}}-M_{\tilde{q}}$ plane.  The dashed lines are the uncorrected bounds, while the solid lines include the QCD corrections.  The region below the lines is excluded.  Going left to right: $\delta_L=\delta_R=0.03,0.1,0.3,1.0$.}
\label{results-equal}
\end{figure}

\begin{figure}[tb]
\centerline{\scalebox{0.5}{
\includegraphics{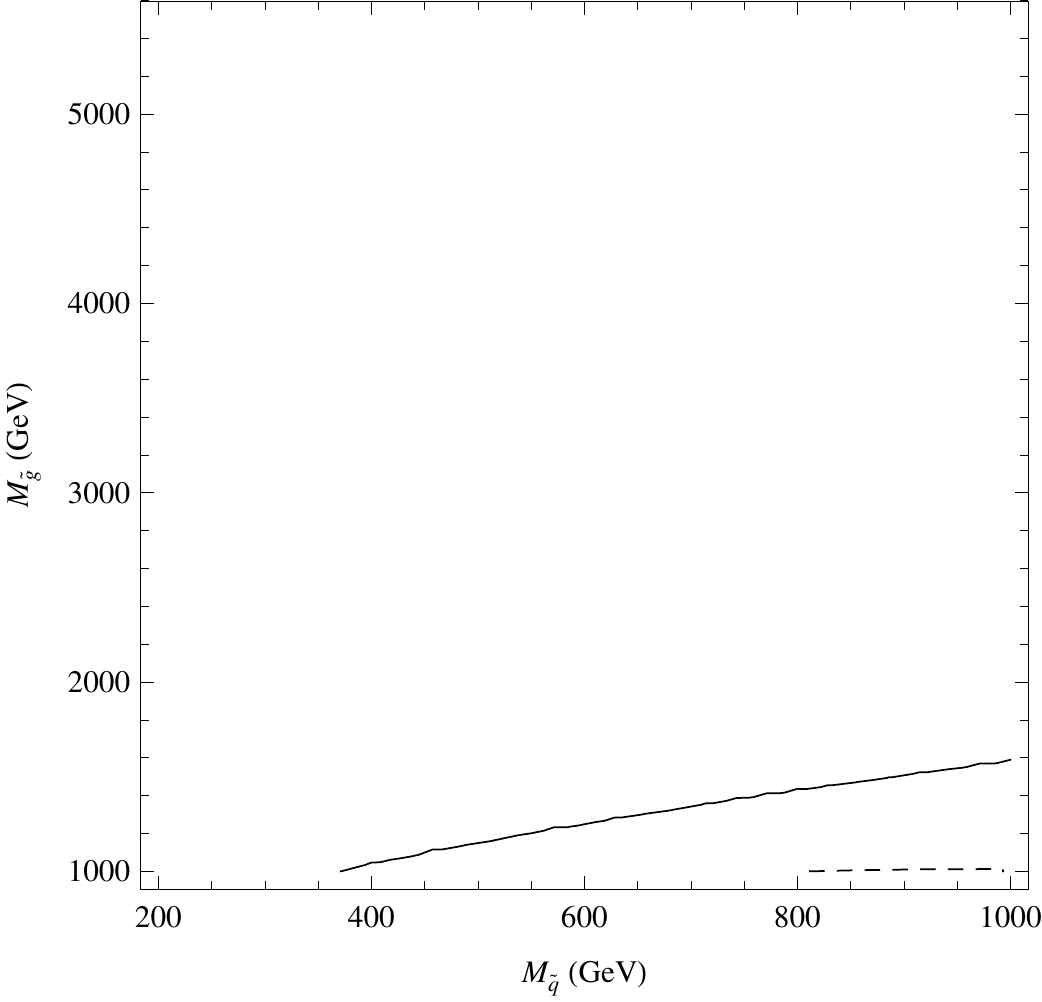}\hspace{0.1cm}
\includegraphics{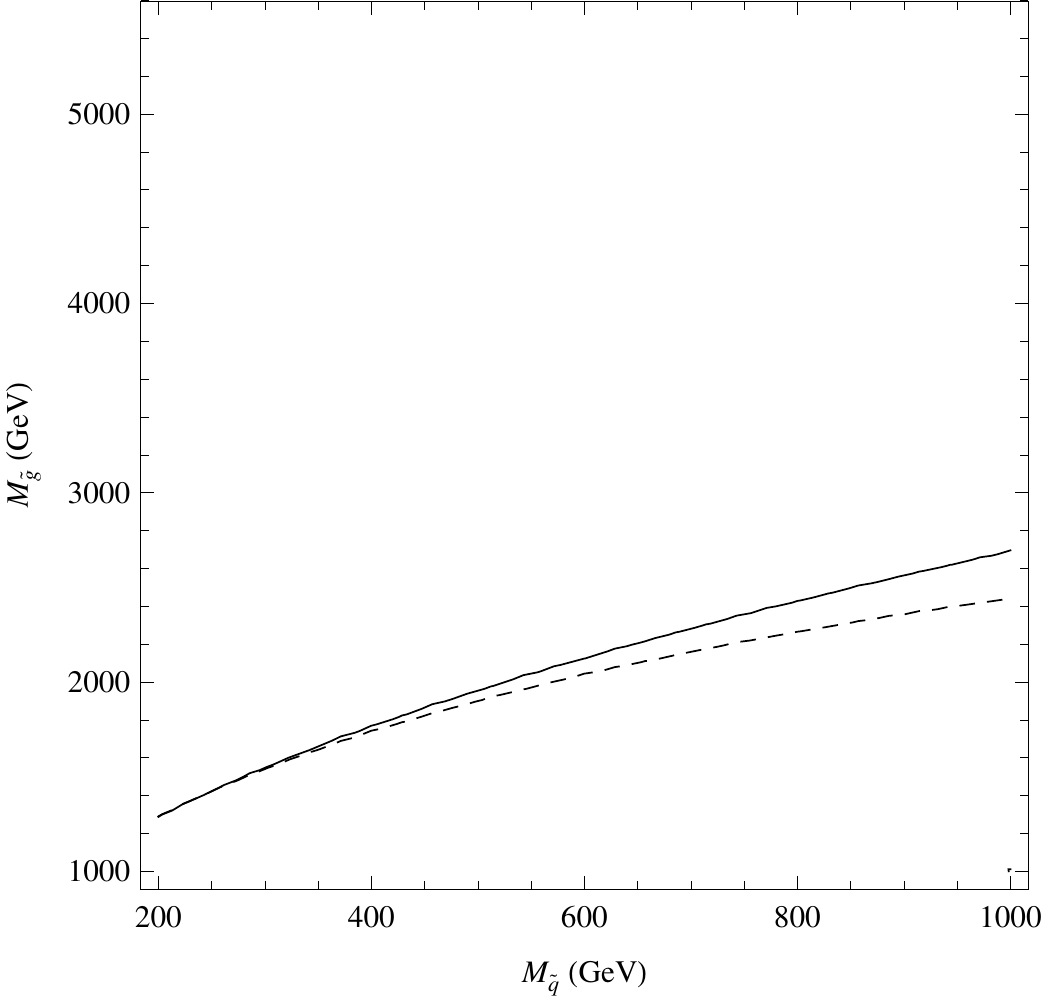}\hspace{0.1cm}
\includegraphics{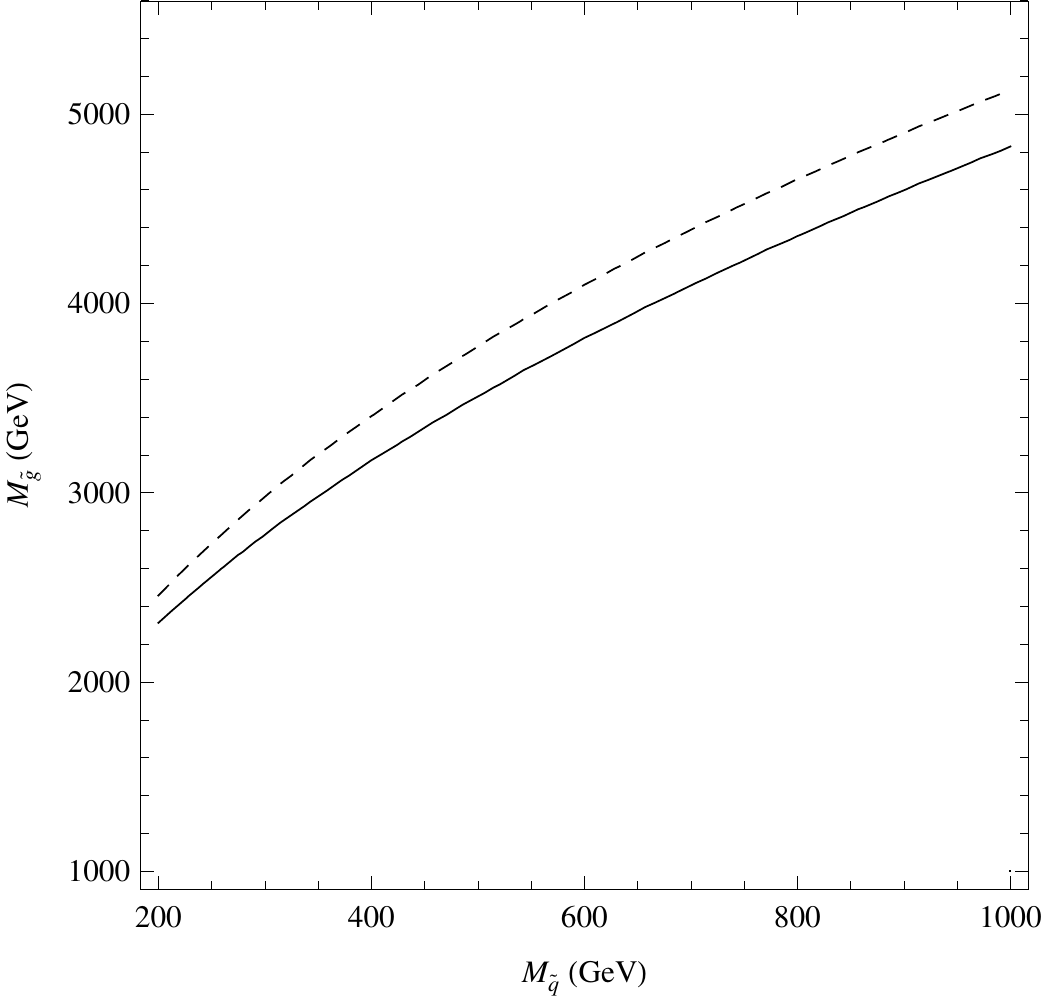}}}
\vspace{-0.2cm}
\caption{Same as Figure \ref{results-equal}, but with $\delta_R=0$.  From left to right: $\delta_L= 0.1,0.3,1.0$.}
\label{results-notequal}
\end{figure}

\subsection{$K-\bar{K}$ Mass Difference}

Now that we have computed the leading contributions to the effective Lagrangian in Equation \ref{Leff} we can proceed to place bounds on the model by looking at the $K-\bar{K}$ mass difference given by
\be
\Delta m_K = 2{\rm Re}\left(\langle K|\mathcal{L}_{\rm eff}|\bar{K}\rangle\right)
\ee
This number is experimentally constrained to be less than $3.48\times 10^{-12}$ MeV.  In what follows we will use the numerical values for the matrix element parameters given in Table \ref{params}.  In addition, we use $\alpha_s(M_Z)=0.118$.

In Figure \ref{results-equal}-\ref{results-notequal} we plot the upper bounds in the $M_{\tilde{g}}-M_{\tilde{q}}$ plane for various values of $\delta_L$ and $\delta_R$.  In Figure \ref{results-equal} we show the plot for a universal value of $\delta$.  It is very clear from these plots that the QCD corrections are very significant, roughly a factor of three stronger.  This is similar to the results of \cite{Bagger:1997gg}.  This somewhat weakens the resolution to the flavor puzzle in \cite{KPW}, but it does not destroy it.  In particular, for $\delta\sim 1$ the model is strongly constrained, but even for $\delta\sim 0.1$ it is still very open.

Of course, for $\delta\sim\mathcal{O}(1)$ our approximations formally break down, since such a large off-diagonal mass term would imply a larger splitting between the $\tilde{d}$ and $\tilde{s}$ squarks.  While we do not expect this to change the bounds significantly, it is worth remembering when trying to do a careful phenomenological analysis of the model.

In Figure \ref{results-notequal} we do the same analysis but with $\delta_R=0$.  In this case only $C_1$ is nonzero.  As is clear from the figure, the QCD corrections do very little to the constraints.  

That we roughly agree with \cite{Bagger:1997gg} is not unexpected.  In both cases, it is the low energy physics below the squark scale that dominates in the running.  You can guess that this is the case since at lower scales (1) $\alpha_s$ is larger and (2) the resummed logs are larger: $\log(m_{\tilde{g}}/m_{\tilde{q}})\ll \log(m_{\tilde{q}}/\mu_{\rm had})$ \cite{KA}.  Nevertheless, is is good to show the full calculation to give quantitative results.

\subsection{CP Violation}

In addition, there is also a nontrivial phase that can enter through the product $\delta_L^{*}\delta_R$, so we can consider the effect of this phase on the CP--violating parameter
\be
\left|\epsilon_K\right|=\frac{{\rm Im}\left(\langle K|\mathcal{L}_{\rm eff}|\bar{K}\rangle\right)}{\sqrt{2}\Delta m_K}
\ee
We can read off the bounds on CP violating phases from Figure \ref{results-equal} as explained in \cite{KPW}.  If we let $\theta\equiv\arg(\delta_L^{*}\delta_R)\ll 1$, we have that $\epsilon_K\propto\theta$.  If we use the example points in \cite{KPW}, $M_{\tilde{g}}=3500$ GeV,  $M_{\tilde{q}}=400$ GeV, and $\delta_L=\delta_R=0.06$ $(0.25)$, and assume that the SUSY contribution saturates the bound\footnote{This is what was done in \cite{KPW}.}, we find that $\theta<9.8$ $(0.57)\times 10^{-3}$.  This is an order of magnitude stronger than the uncorrected results and suggests that the phase might have to be unnaturally small in order to avoid experimental constraints.

\section{Conclusion}
In this paper, we have computed the leading-log QCD corrections to the model discussed in \cite{KPW}.  It was realized then that an $R-$symmetry built into a SUSY model with Dirac gauginos naturally avoids the flavor puzzle found in typical SUSY models.  We find that QCD corrections can significantly affect this result by nearly a factor of 3, similar to the usual SUSY story in \cite{Bagger:1997gg}.  The model is still significantly less fine tuned, with $\delta\sim 0.1$ allowed in a typical models, although in the universal case, CP violation can still be very constraining.  It is clear that QCD corrections are important in general and must be included when studying FCNC processes in R-symmetric supersymmetric models.

\appendix
\section{Anomalous Dimension Matrix in the Intermediate Theory}

\begin{figure}[tb]
\centerline{\scalebox{0.7}{\includegraphics{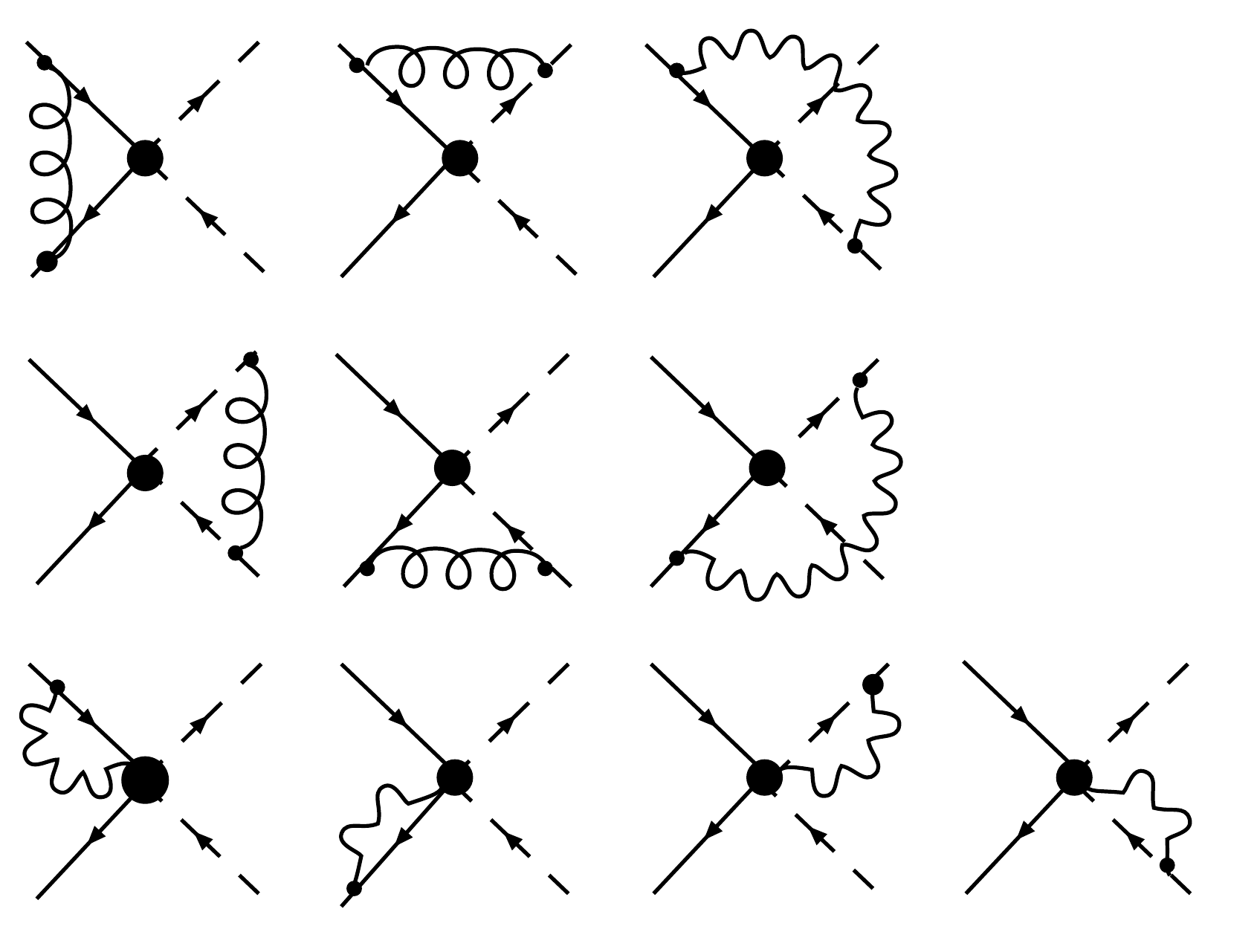}}}
\vspace{-0.2cm}
\caption{QCD loops that contribute to the leading order counterterms to Class A operators.  The Class B operators are similar.}
\label{running}
\end{figure}

The intermediate theory is described up to flavor structure by the six dimension-6 operators given in Equations \ref{Qop1}-\ref{Qop6}.  To compute the QCD anomalous dimension matrix, one must evaluate the diagrams in Figure \ref{running} for each of these operators.  The Class A operators $\mathcal{Q}^{1,2}_{1,3,5}$ have a different color structure than the corresponding Class B operators $\mathcal{Q}^{3,4}_{1,3,5}$ and will therefore run differently.  However, the operators $\mathcal{Q}^{1,2,3,4}_{2,4,6}$ have the same color structure and therefore both classes run the same way; this has been confirmed by direct calculation.  The final result is a $6\times 6$ matrix for each class:
\be
\gamma_A=\frac{\alpha_s}{2\pi}\left(\bay{cccccc}
\vspace{0.1in}-\frac{15}{4} & -\frac{11}{4} & 0 & 0 & -\frac{15}{4} & +\frac{5}{4} \\
\vspace{0.1in}-\frac{7}{2} & -\frac{3}{2} & 0 & 0 & 0 & 0 \\
\vspace{0.1in} -\frac{14}{3} & -2 & +\frac{8}{3} & 0 & -6 & +2 \\ 
\vspace{0.1in}-2 & -\frac{14}{3} & -\frac{3}{4} & +\frac{59}{12} & 0 & 0 \\
\vspace{0.1in}-\frac{91}{12} & -\frac{13}{4} & 0 & 0 & +\frac{5}{12} & +\frac{3}{4} \\
\vspace{0.1in}-\frac{13}{4} & -\frac{91}{12} & 0 & 0 & -\frac{3}{4} & +\frac{59}{12}
\eay\right)
\qquad
\gamma_B=\frac{\alpha_s}{2\pi}\left(\bay{cccccc}
\vspace{0.1in}-\frac{39}{4} & -\frac{3}{4} & 0 & 0 & +\frac{15}{4} & -\frac{5}{4} \\
\vspace{0.1in}-\frac{7}{2} & -\frac{3}{2} & 0 & 0 & 0 & 0 \\
\vspace{0.1in} -\frac{14}{3} & -2 & -\frac{67}{12} & +\frac{11}{4} & 0 & 0 \\ 
\vspace{0.1in}-2 & -\frac{14}{3} & -\frac{3}{4} & +\frac{59}{12} & 0 & 0 \\
\vspace{0.1in}-\frac{23}{6} & -\frac{9}{2} & -6 & +2 & -\frac{10}{3} & +2 \\
\vspace{0.1in}-\frac{13}{4} & -\frac{91}{12} & 0 & 0 & -\frac{3}{4} & +\frac{59}{12}
\eay\right)
\ee

\begin{acknowledgments}
A.E.B. and S.P.N. were supported by the Ontario Premier's Research Excellence Award and the Natural Sciences and Engineering Research Council of Canada. The authors would like to thank Kaustubh Agashe, David E. Kaplan, Erich Poppitz and Neal Weiner for useful discussions, and Graham Kribs for a careful reading of the manuscript.  We are also grateful to the authors of \cite{KPW} for sharing their code, which was used for comparative purposes.
\end{acknowledgments}

\newpage

%

\begin{thebibliography}{99}
\bibitem{KA} We would like to thank Kaustubh Agashe for pointing this out.
\end{thebibliography}
\end{document}